\documentclass[aps,pre,twocolumn,showpacs,superscriptaddress]{revtex4}
\usepackage{epsfig}
\usepackage{times}
\begin{document}

\title{Reciprocity-based cooperative phalanx maintained by overconfident players}

\author{Attila Szolnoki}
\email{szolnoki.attila@energia.mta.hu}
\affiliation{Institute of Technical Physics and Materials Science, Centre for Energy Research, Hungarian Academy of Sciences, P.O. Box 49, H-1525 Budapest, Hungary}

\author{Xiaojie Chen}
\email{xiaojiechen@uestc.edu.cn}
\affiliation{School of Mathematical Sciences, University of Electronic Science and Technology of China, Chengdu 611731, China}

\begin{abstract}
According to the evolutionary game theory principle, a strategy representing a higher payoff can spread among competitors. But there are cases when a player consistently overestimates or underestimates her own payoff, which undermines proper comparison. Interestingly, both underconfident and overconfident individuals are capable of elevating the cooperation level significantly. While former players stimulate a local coordination of strategies, the presence of overconfident individuals enhances the spatial reciprocity mechanism. In both cases the propagations of competing strategies are influenced in a biased way resulting in a cooperation supporting environment. These effects are strongly related to the nonlinear character of invasion probabilities which is a common and frequently observed feature of microscopic dynamics.
\end{abstract}


\maketitle

\section{Introduction}
The basic idea of evolutionary game theory is to consider payoff as fitness, and higher payoff in the game is translated into reproductive success. Hence, due to natural selection, more successful strategies reproduce faster while those strategies which are less successful become extinct \cite{nowak_06}. This microscopic dynamic assumes an accurate comparison of payoff values which help competitors to navigate toward a better evolutionary outcome. 

Interestingly, however, we can observe examples in real life situations when individuals tend to misinterpret their own payoff values perpetually hence the estimation of payoff difference can be easily misleading. Overconfidence, believing more about themselves than they are in reality, could be a source of biased belief \cite{trivers_11}. But underconfident players, who believe less about themselves, make decisions based on incorrect perception, too. Indeed, the possible evolutionary advantage of overconfidence in resource competition games has already been revealed by previous works \cite{johnson_ddp_n11, li_k_srep16}. The main conclusion of these works was that overconfidence could be beneficial because it encourages individuals to claim resources they could not otherwise win. Secondly, overconfidence keeps these competitors from walking away from conflicts they would probably win. 

But what if {\it all} players are overconfident or {\it all} players are underconfident when they estimate their own achievement? One may expect that if all members commit the same error of perception then there is no relevant change from the viewpoint of evolutionary dynamics. In this work we focus on this question by considering the fundamental problem of cooperation \cite{sigmund_10}. Here cooperator and defector strategies compete and to defect would always provide a higher individual income against a cooperator, but mutual cooperation would offer the optimal income for the whole community. In the last decades several cooperator supporting mechanisms were identified \cite{nowak_s06}, including reward \cite{jimenez_jtb08, szolnoki_epl10, hilbe_prsb10, sasaki_bl14, wu_y_srep17} or punishment \cite{fehr_n02,helbing_ploscb10, nakamaru_jtb06, han_ab16, szolnoki_prx17, takesue_epl18}, population heterogeneity \cite{santos_prl05,perc_pre08,santos_n08}, player's mobility \cite{chen_w_pa16,cong_r_srep17}, conformity \cite{szolnoki_rsif15, yang_hx_csf17} and tolerance \cite{szolnoki_pre15, riolo_n01}, which could be helpful to avoid the tragedy of the common state when everyone chooses the tempting defection \cite{hardin_g_s68}. 

In this work we will not assume any sophisticated environmental feedback mechanism \cite{alonso_jsm06,chen_xj_pre09b,szolnoki_epl17} or demanding cognitive skill about players \cite{press_pnas12, hilbe_pone13b, mobilia_pre12, han_srep15, hilbe_pnas13, mobilia_csf13}, but only explore the plain consequence of perception error collectively made by group members. To reveal the interaction between strategies and individual skills we consider a coevolutionary model \cite{perc_bs10,stivala_pre16,richter_bs17,khoo_srep18,takesue_epl17} where players may not only imitate a more successful strategy but also adopt the way to consider individual achievement when making decisions. In particular, besides individuals who are performing unbiased estimation of their payoff values we also assume the initial presence of over- and underconfident players and monitor the coevolutionary process. Interestingly, being overconfident not only ensures individual advantage but could also be beneficial for the whole community if everyone follows the same trait. Furthermore, a higher cooperation level can also be reached when all members of the population are underconfident regarding their own success. These observations can be explained dynamically by a modified microscopic process which has a biased consequence on strategy propagations. 

The organization of this paper is as follows. We first present the definition of our model in the next section. We then proceed with the presentation of our main results and their explanations. This is followed by our conclusions and a discussion of their implications in the last section. 

\section{Coevolution of perception and player strategies}

Starting from the traditional prisoner's dilemma game we assume that unconditional cooperator and defector players are distributed on a graph. For simplicity we use a square lattice interaction graph, but we stress that our observations remain valid for other types of interaction networks.

To capture the essence of a social conflict we adopt the simplified parametrization of weak prisoner's dilemma game \cite{nowak_n92b} where the only parameter is the temptation to defect $T$, while reward for mutual cooperation provides $R = 1$ payoff. The punishment $P$ for mutual defection as well as the sucker's payoff $S$ of a cooperator facing a defector are equal to 0.

The evolution of the competing strategies is performed in accordance with the following elementary steps. First, a randomly selected player $x$ acquires its payoff $\Pi_x$ by playing the game with all its $k_x$ neighbors. Next, a randomly chosen neighbor of $x$, denoted by $y$, also acquires its payoff $\Pi_y$ by playing the game with all its $k_y$ neighbors. Last, player $x$ adopts the $s_y$ strategy of player $y$ with a probability 
\begin{equation}
\Gamma(s_y \to s_x)= 1/ \{1+\exp[(\Pi_x-\Pi_y) /K]\}\,\,,
\label{fermi}
\end{equation}
where $K$ denotes the amplitude of noise that quantifies the uncertainty of strategy adoptions \cite{szabo_pre98,vukov_pre06}.

The only difference from the traditional model is we assume that players may have different levels of self-deception when they evaluate their own payoff values for pairwise comparison. For simplicity we establish three classes for self-deception, which are underconfident ($u$), normal ($n$) and overconfident ($o$) players. Traditionally, when normal players calculate the imitation probability they apply an unbiased (or accurate) payoff value for their own achievement.
An overconfident player $x$, however, believes more about her own achievement than its proper $\Pi_x$ value. Consequently, she will use an enhanced $\Pi^\prime_x = \Pi_x (1+\alpha)$ payoff value to calculate the imitation probability. Here parameter $\alpha$ describes the level of overconfidence.
Similarly, an underconfident player $x$ underestimates her own achievement and uses a reduced $\Pi^\prime_x = \Pi_x (1-\alpha)$ value when imitation probability is calculated. For simplicity we use the same parameter to characterize the degree of biased self-deception to both directions. Notably, the self-deception level can also be adopted via a learning step with the same probability but the latter option is only considered when players have different strategies. Otherwise, the overconfident state would always enjoy an artificial advantage over other states even within a homogeneous-strategy domain. Nevertheless, we note that the final cooperation level remains intact if we allow the adoption of confidence level between players with identical strategies. To summarize the microscopic dynamics of our model a personal strategy and confidence level can be adopted independently, but using the same adoption probability which is based on the payoff difference of source and target players. In other words, it can happen that only a confidence level is adopted while the strategy of target player remains unchanged, or only strategy invasion happens, or both features are adopted simultaneously.

Technically we consider a six-strategy model where besides traditional or normal $C_n$ and $D_n$ players we also have overconfident $C_o$ and $D_o$ players and underconfident $C_u$ and $D_u$ competitors. It is important to stress that the increase (decrease) of payoff for overconfident (underconfident) players is conceptually different from the general perception error that is captured via the noise parameter $K$. While perception error may emerge toward both directions and an ordinary player sometimes may overestimate or underestimate payoff values, but overconfident (underconfident) players tend to use biased values always into one direction. The key parameters of our coevolutionary model is the temptation $T$ value which characterizes the dilemma strength and the $\alpha$ value which describes how biased the over- and underconfident players are.

Monte Carlo simulations of the game are carried out comprising the described coevolutionary steps. Each Monte Carlo step ($MCS$) gives a chance for every player to adopt the strategy and/or self-deception level of a randomly chosen neighbor once on average. During the evolutionary process we monitor both strategies and the fractions of different self-deception levels. When regular interaction graphs were used (such as square lattice or kagome lattice) the linear size of the system was between $L=400$ and $L=4000$. The typical time to reach a stationary state was 50000 $MCS$s, and we averaged the stationary values over another 10000 steps. For heterogeneous graphs, like random or scale-free graphs, we used $N=5000$ nodes and generated 1000 independent graphs to average the obtained values for the requested accuracy. As already noted, in the following we present the details of results obtained mostly on a square grid, but conceptually similar results can be reached for other interaction graphs.

\section{Results}

Before presenting our results for structured populations we note that in a well-mixed, unstructured population where players have random temporary connections the introduction of biased confidence levels has no particular consequence. More precisely, defector players always prevail for any $T>1$ value in agreement with the classical model \cite{sigmund_10}. Therefore a spatially structured population, which is a rather realistic assumption in several cases, is a fundamental condition for the results discussed below. As a general observation, overconfident players will always prevail in the whole population if we wait long enough. This behavior, which agrees with the prediction obtained for the resource competition game \cite{johnson_ddp_n11}, is not really surprising because these players are reluctant to adopt the state of other competitors while normal and especially underconfident players can do it more easily. But our principal interest is to explore how the presence of players with biased self-deception may influence the cooperation level. This point could be specially interesting in the situation when both cooperators and defectors are overconfident and overestimate their own achievements simultaneously.

\begin{figure}
\centerline{\epsfig{file=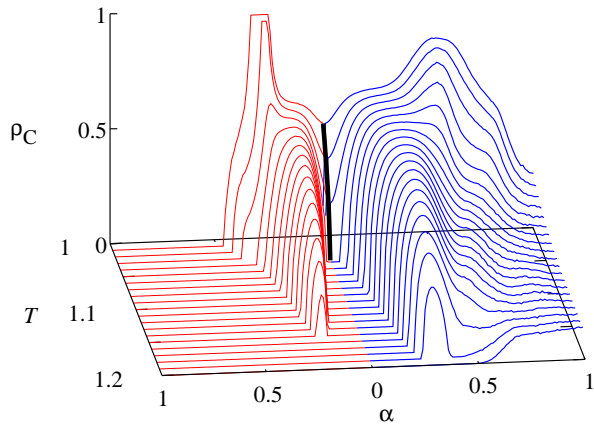,width=8.0cm}}
\caption{Fraction of cooperators on $T-\alpha$ plane for a square lattice at $K=0.1$ in the case when only underconfident players are present (left side, red lines) and in the case when only overconfident individuals are present in the population (right side, blue lines). The latter is also the evolutionary outcome of the general model when players with different confidence levels compete for space. The cooperation level for the normal system is marked by a thick black line at $\alpha=0$.}
\label{sqr} 
\end{figure}

The answer to this question can be found on the right-hand side of Fig.~\ref{sqr} where we plotted the general cooperation level on the $T - \alpha$ plane. This surface suggests that by using an intermediate $\alpha$ value a significantly high cooperation level can be reached even at a large temptation value where a normal system would terminate onto a full defector state. Evidently, if $\alpha$ is too large then players would evaluate their payoff values too high, which would result in a frozen state (not shown in Fig.~\ref{sqr}). But staying at a moderate $\alpha$ the full collapse of the cooperator state can be avoided, which means that a certain level of overconfidence of all members could be useful for the whole community.

Interestingly, not only overconfident but also underconfident players can be useful for the whole community. If we assume a uniform population where all players are underconfident and underestimate their own payoff values then the cooperation level can also be elevated comparing to the normal system where every player estimates payoff values accurately. This observation is summarized on the left-hand side of Fig.~\ref{sqr} where we again plotted cooperation level on the $T - \alpha$ plane, which is the fraction of $C_u$ players in this case. As for the overcondfident case, here there is again an optimal intermediate $\alpha$ value which provides the highest cooperation level.

\begin{figure}
\centerline{\epsfig{file=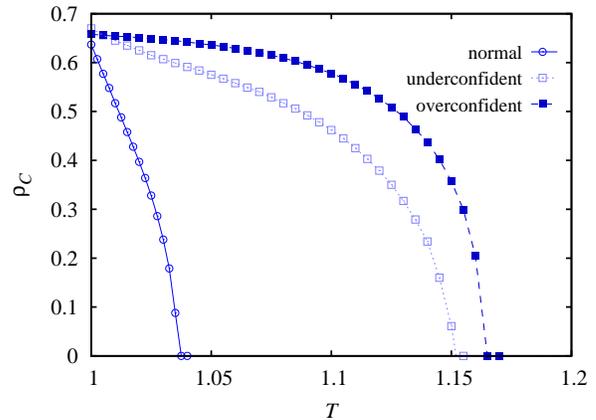,width=8.5cm}}
\caption{Comparison of cooperation levels for uniform models when players with a single-type confidence level are present in dependence on temptation value at  $\alpha=0.2$. In all cases players are staged on a square lattice ($L=400$) at $K=0.1$.}
\label{A20} 
\end{figure}

In Fig.~\ref{A20} we compare the results of uniform populations obtained at a fixed $\alpha$ value. These plots highlight that the positive consequence of biased self-deception is more visible at high temptation values which would normally ensure a clear advantage for defector players. Furthermore, an overconfident population can do even better than an underconfident population.

\begin{figure}
\centerline{\epsfig{file=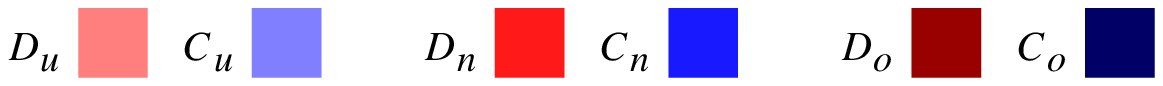,width=5.8cm}}
\centerline{\epsfig{file=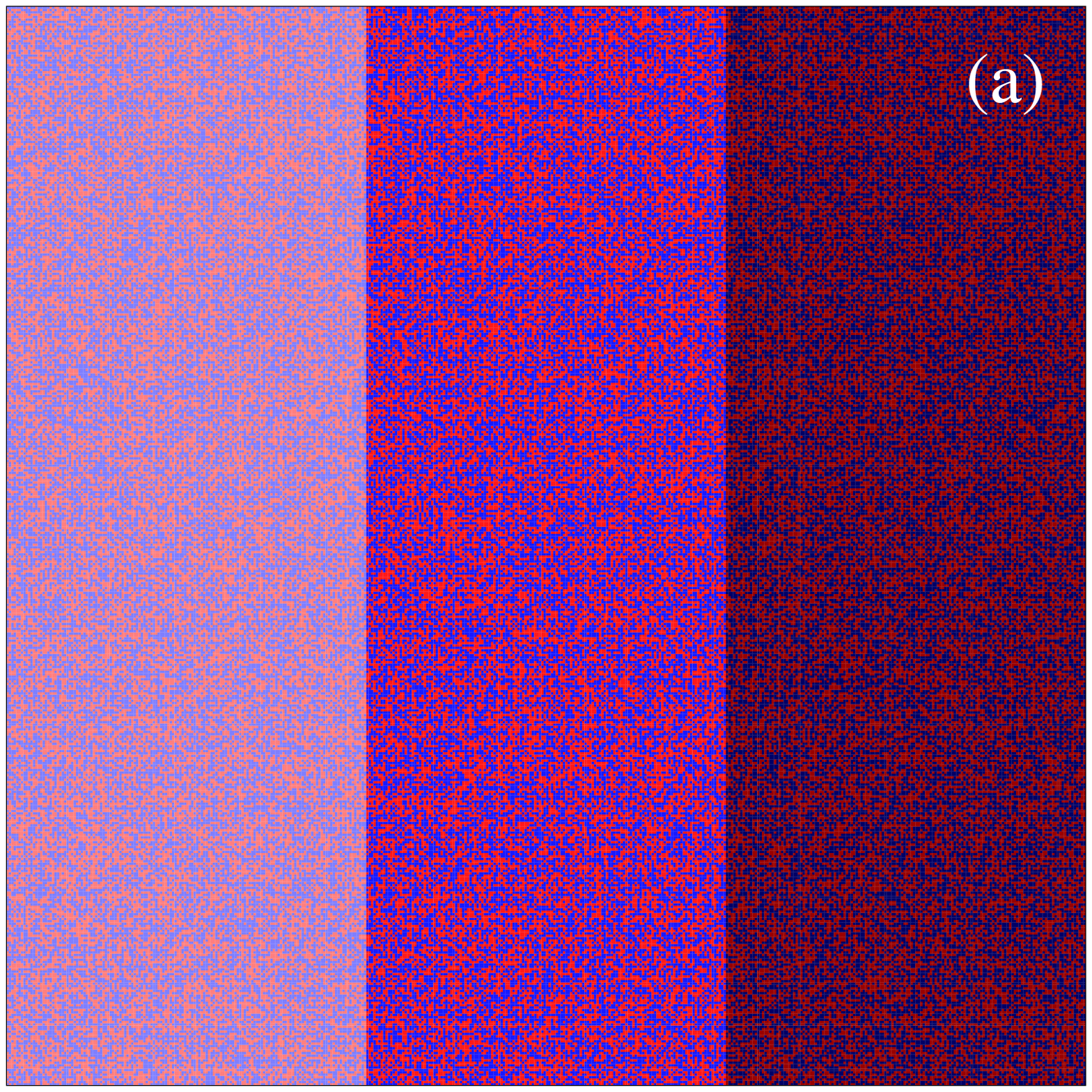,width=5.1cm}}
\centerline{\epsfig{file=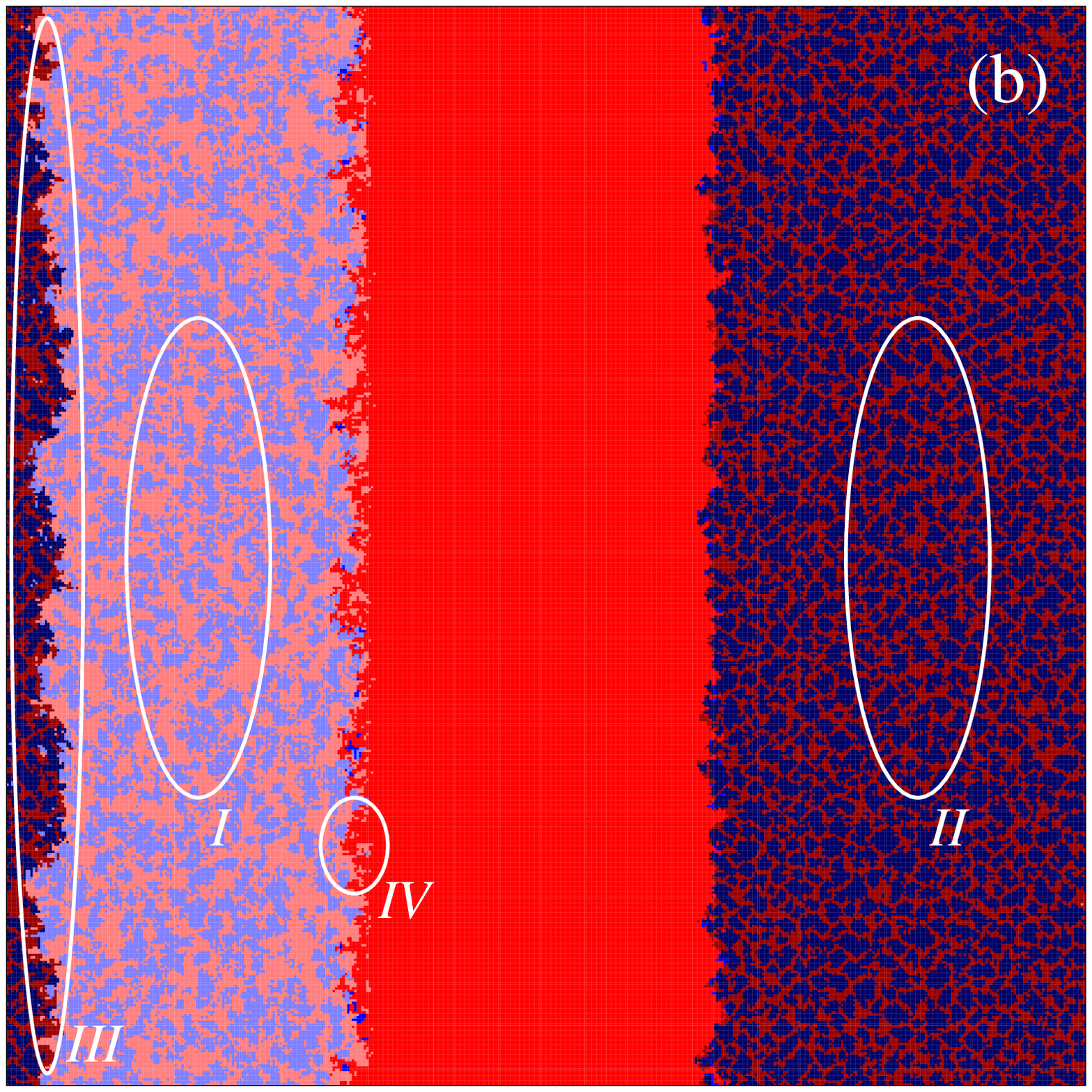,width=5.1cm}}
\centerline{\epsfig{file=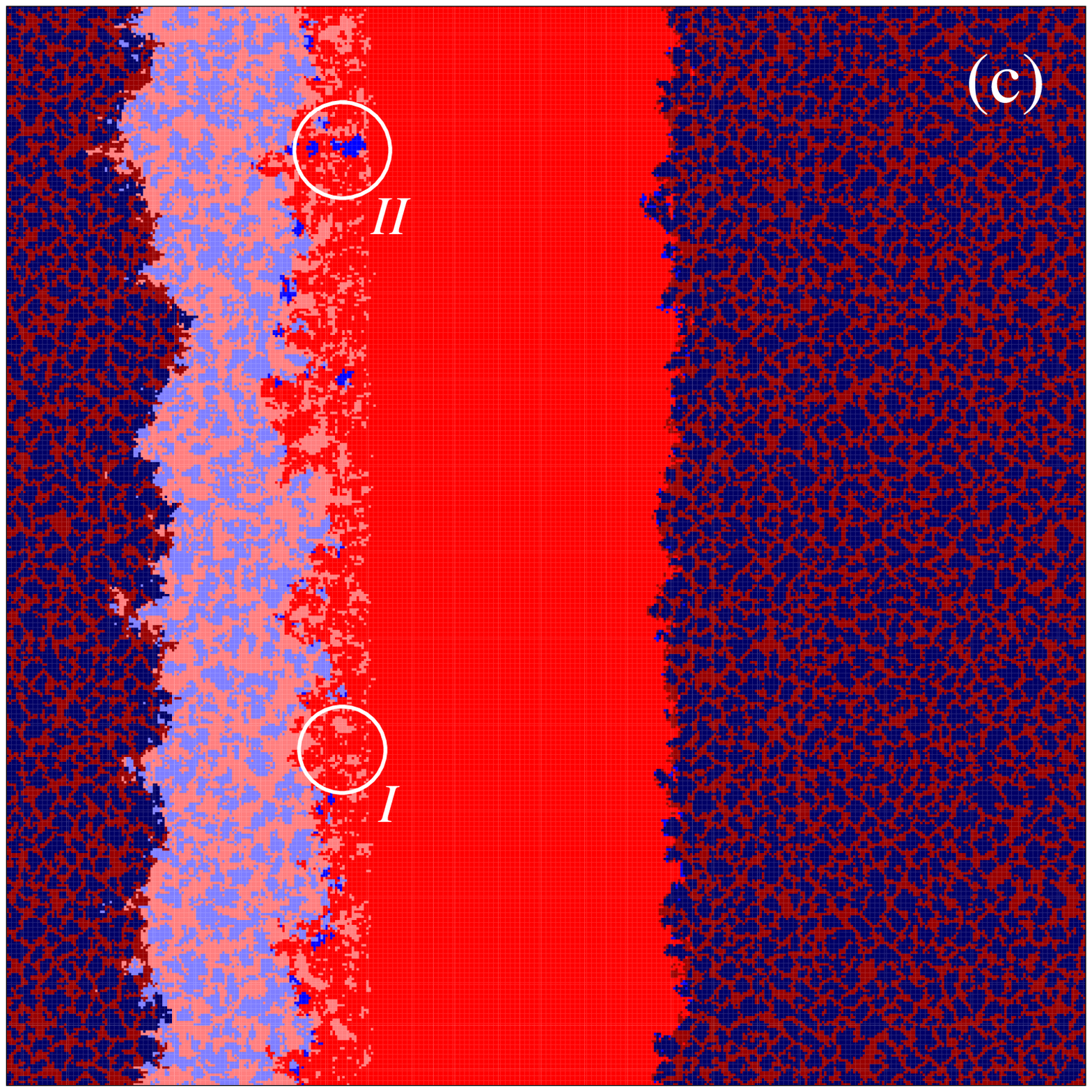,width=5.1cm}}
\centerline{\epsfig{file=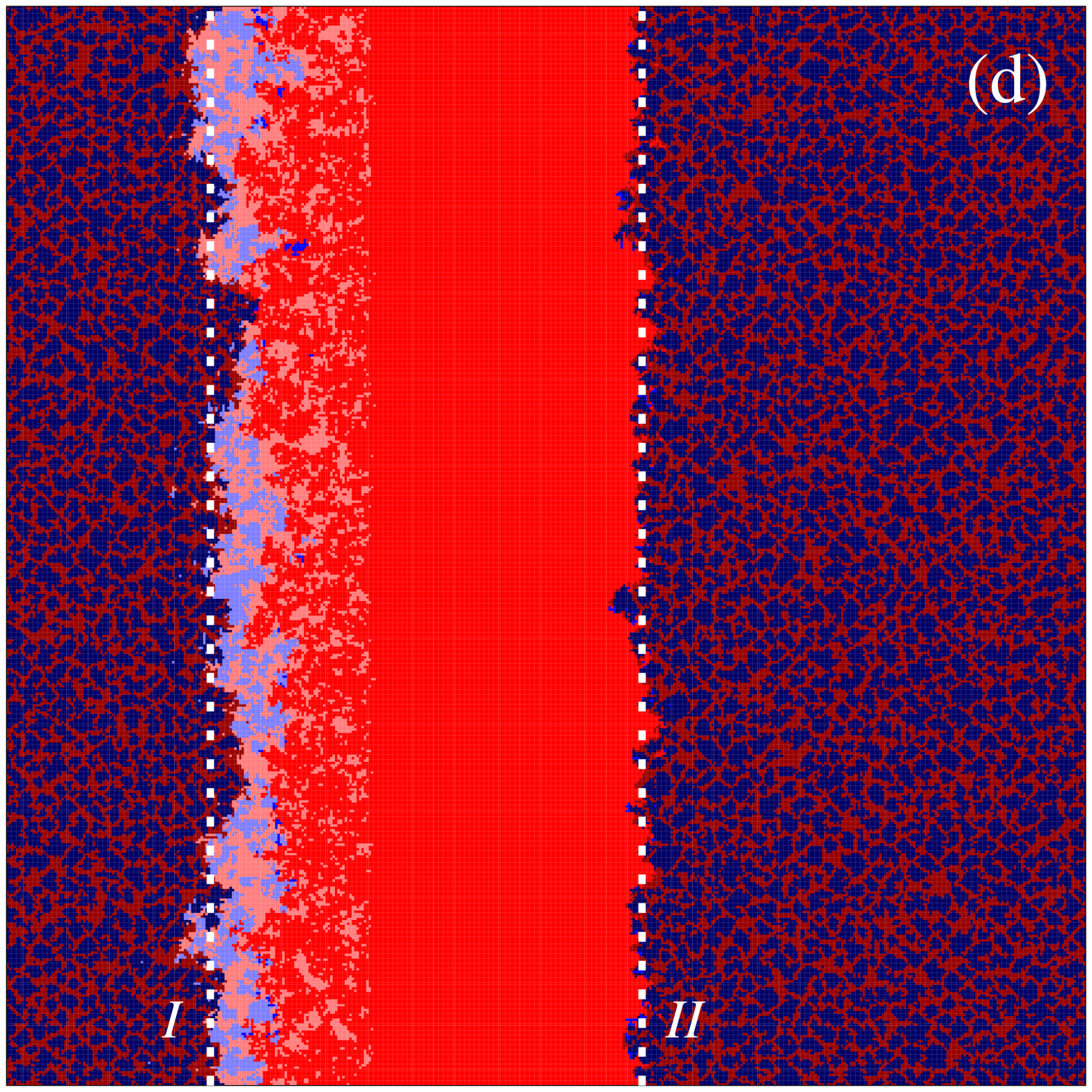,width=5.1cm}}
\caption{Competition of possible solutions at $T=1.1, \alpha=0.2$ on a square lattice with $L=450$ linear size. Different shade of blue and red colors denote cooperator and defector players with different self-deception levels as indicated by the legend on the top. 
Further details are given in the main text. Snapshots were taken at 0, 200, 600, and 850 $MCS$s.}
\label{snapshots} 
\end{figure}

To collect deeper insights into the typical microscopic mechanisms responsible for the coevolutionary process, we present characteristic snapshots of evolution started from a prepared initial state where all available states are present. The whole evolution can be monitored in the animation we provided \cite{animation}
but the milestones of pattern formations are described in the following.
Figure~\ref{snapshots}~(a) shows the starting state where players with different self-deception levels are distributed separately. In particular, underconfident cooperator and defector players are arranged randomly in the left third of the space. Overconfident players are initially distributed on the right third, while normal players with unbiased self-deception are in the center third. When evolution is launched then sub-solutions emerge locally. More precisely, as shown in Fig.~\ref{snapshots}~(b), the applied large temptation value prevents normal $C_n$ cooperators to survive in the sea of normal $D_n$ defectors. In biased populations, however, cooperators survive. As the area marked by $``I"$ illustrates, $C_u$ players coexist with $D_u$ defectors, and similarly $C_o$ cooperators form a solution with $D_o$ defectors in the region marked by $``II"$. There is a visible difference between these two solutions, which will have a greater importance as discussed below. In particular, $C_o$ players form compact domains in the sea of $D_o$ players while the domains of $C_u$ players are more irregular.

Due to periodic boundary conditions, overconfident players can interact directly with underconfident players, shown by area $``III"$, and the former solution prevails against the latter. The stability of $C_u + D_u$ solutions is also jeopardized by normal players because $D_n$ invades the territory of $C_u$, as shown by $``IV"$ in Fig.~\ref{snapshots}~(b). Because of the applied microscopic dynamic, which allows the adoption of self-deception level only between players with different strategies, $D_n$ and $D_u$ states would coexist, as illustrated by $``I"$ in Fig.~\ref{snapshots}~(c). This mixture, however, is not stable because some $C_u$ players adopt the self-deception level from $D_n$ neighbor and the emerging $C_n$ state can easily spread in a $D_u$ domain and sweep them out completely. This process is shown by $``II"$ in panel~(c). The triumph of $C_n$, however is just temporary because neighboring $D_n$ players beat them, as it is explained previously. Summing up, albeit $D_n$ and $D_u$ are neutral, but the former beats the latter indirectly with the help of $C_n$ players, who directly invades $D_u$ and after becomes the prey of $D_n$. This pattern formation resembles ``the Moor has done his duty, the Moor may go" effect that emerges in several multi-state ecological systems \cite{szolnoki_pre11b, danku_epl18}.

As we already noted at the beginning of this section, overconfident players invade the whole populations. This invasion can be seen clearly in Fig.~\ref{snapshots}~(d) where we marked both propagation fronts of this domain. The front marked by $``I"$ denotes the irregular, but fast propagating overconfident $\to$ underconfident transition. Here both $C_o \to D_u$ and $D_o \to C_u$ elementary steps assist the propagation.
The front marked by $``II"$ separating normal and overconfident players is more regular, but propagates much slower. Here only $C_o$ players can invade $D_n$ territory first which is followed by the some invasion between $D_o$ and $C_o$ players, which establishes the stable coexistence of the latter states. Finally, not shown in Fig.~\ref{snapshots}, only these two types of players remain alive.

\begin{figure}
\centerline{\epsfig{file=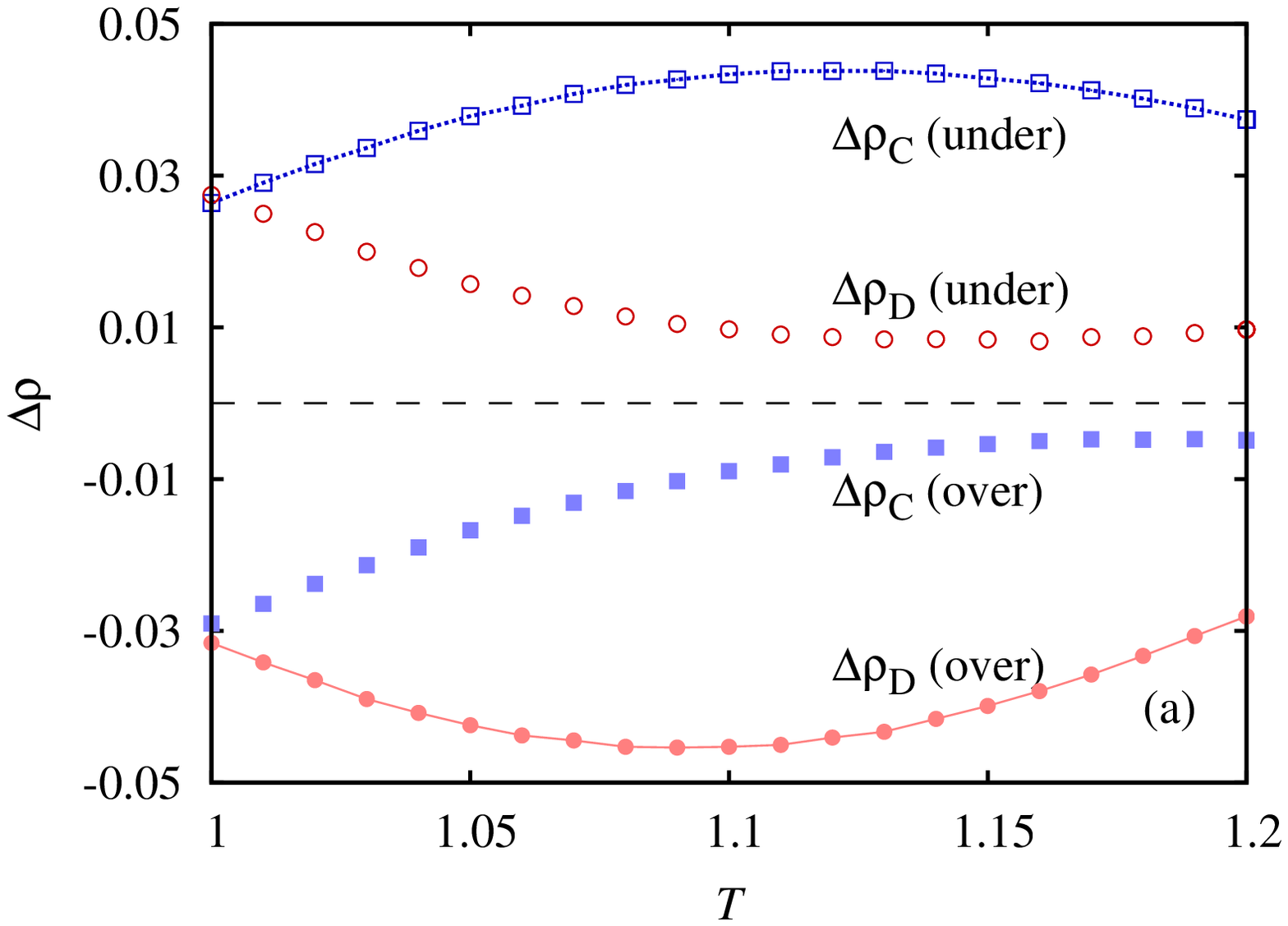,width=8.5cm}}
\centerline{\epsfig{file=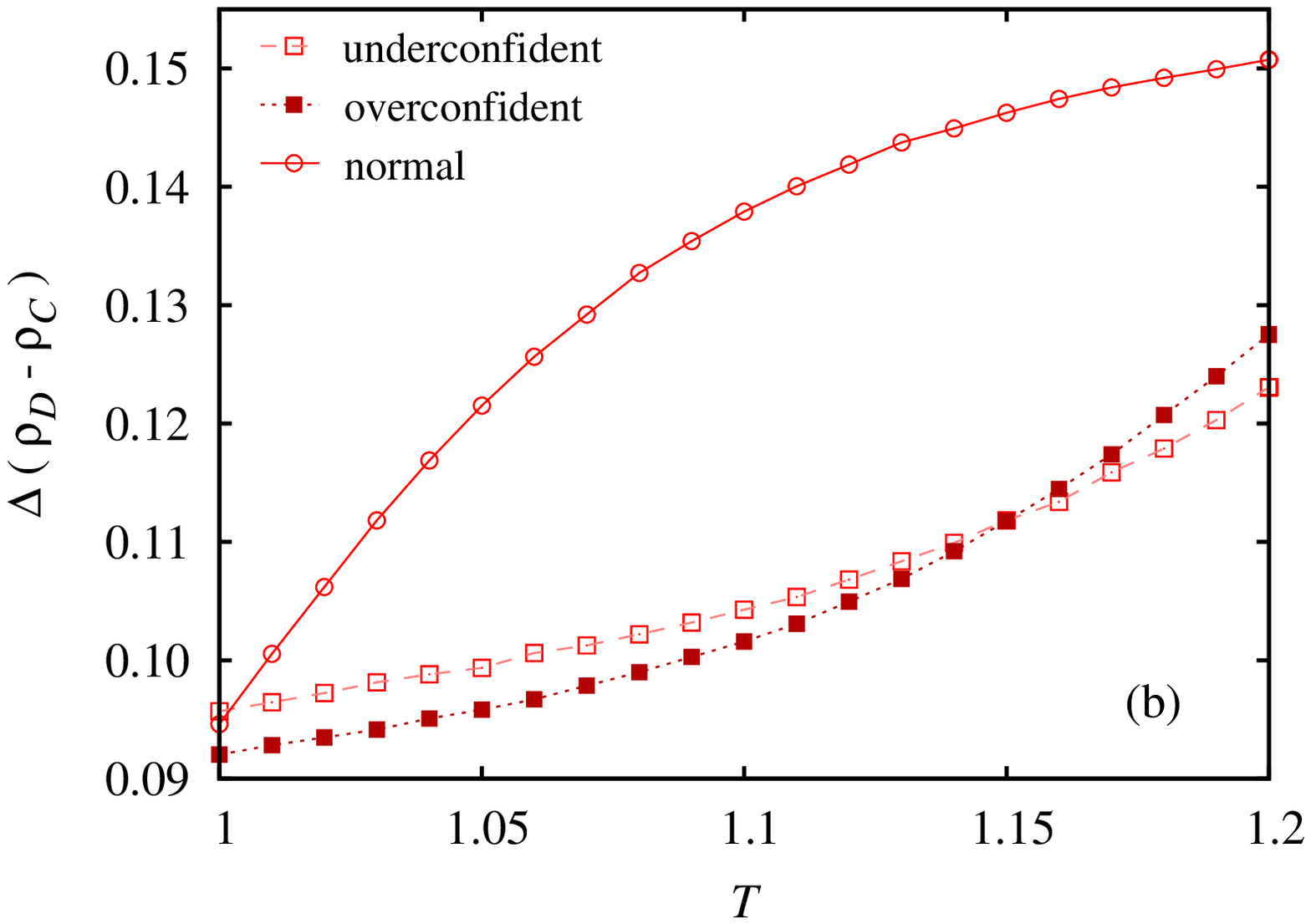,width=8.5cm}}
\caption{Changes of invasion speeds detected via the development of strategy concentrations when uniform systems are launched from a random initial state on a $L=4000$ square lattice at $\alpha=0.2$. In panel~(a) $\Delta \rho_C$~(under) and $\Delta \rho_D$~(under) show how the fractions of cooperators and defectors change during elementary invasion steps when we replace normal players by underconfident individuals. Similarly, $\Delta \rho_C$~(over) and $\Delta \rho_D$~(over) denote how the fraction of cooperators and defectors vary when we replace normal players by overconfident competitors. Symbols connected by lines mark the leading mechanisms for both cases which are responsible for the cooperation supporting effects summarized in Fig.~\ref{sqr}.
Panel~(b) shows the resulting relative changes of strategies for all uniform models. This comparison suggests that the advantage of defector strategy weakens significantly for both underconfident and overconfident populations.}
\label{inv} 
\end{figure}

To understand why biased populations support cooperation, it is instructive to analyze the propagation processes within uniform systems where players share the same self-deception level. For this purpose we compare the strategy invasions in three different uniform systems by using the same $T, \alpha$ parameter values when evolution is launched from a random initial state. As expected, a random mixture of strategies always supports defector invasion better in the early stage, but its intensity could be different for populations with different self-deception levels. The simplest way to quantify the intensity of invasion is to measure how the fractions of strategies change in time due to elementary invasion steps. For proper comparison we measure how the success of specific invasion steps change when we change the self-deception level for all players. Considering a normal, unbiased system as a reference, Fig.~\ref{inv}(a) shows the change for all invasions when we apply biased populations.
Here $\Delta \rho_C$ shows how the successful invasion steps increasing the cooperation level change if we replace an unbiased system by a biased model. Similarly $\Delta \rho_D$ denotes how the frequency of successful defector invasion steps varies when we change a normal system to a biased one. As expected, the strategy invasions are more intensive for an underconfident population, hence the changes of successful invasion steps are positive for both strategies. But the increment for cooperator strategy is larger than for defector strategy. It simply means that the cooperator invasion in a normal system is so weak that the general increment of adoption skill for underconfident players provides a significant support for $C$ strategy. On the other hand, the invasion success of defection is so strong in a normal system that it does not give relevant additional support for $D$ strategy when we use underconfident players who adopt neighboring strategies more easily.

Also in agreement with our expectation the general strategy invasion is reduced for the overconfident population, hence the change is negative compared to the normal system. But again, the levels of change for different strategies are strikingly different. While the cooperator invasion decreases just slightly, the decline of defector invasion is significant when we replace a normal system by an overconfident population. In other words, defectors loose more when we lower the general adoption capacity because their success in a normal system is significant while cooperator players are rather unsuccessful.

While Fig.~\ref{inv}~(a) shows the change of successful strategy invasions compared to a normal system, panel~(b) of Fig.~\ref{inv} shows the sum of defector and cooperator invasions for all uniform systems. As expected, defectors are more successful for all $T$ values in all cases, but the success of defectors is significantly weakened for both biased systems compared to the normal system which is also plotted here. Albeit the evolutionary consequences are similar, but their explanations are different. When underconfident players are used, their general willingness to change strategy moves the evolutionary dynamics toward a more neutral direction, hence the individual advantage of defection is less straightforward. Consequently, the microscopic dynamics that are less deterministic can be detected from the irregular, noisy patterns of domains we already noted in Fig.~\ref{snapshots}~(b). In the presence of overconfident players all microscopic changes are suppressed in general, but cooperators can benefit more from this fact. Indeed, network reciprocity is strengthened and the phalanx of $C$ becomes more robust which is hardly broken by defectors even for a significantly higher temptation. Therefore the borders of cooperator domains become smooth, and the $C$ domains are more compact, as illustrated in Fig.~\ref{snapshots}~(b). This observation fits nicely into the general expectation that a microscopic rule which strengthens surface tension and smooth separating domain walls could be beneficial for the evolution of cooperation \cite{szolnoki_srep12,szolnoki_pre12,szolnoki_rsif15}.

\begin{figure}
\centerline{\epsfig{file=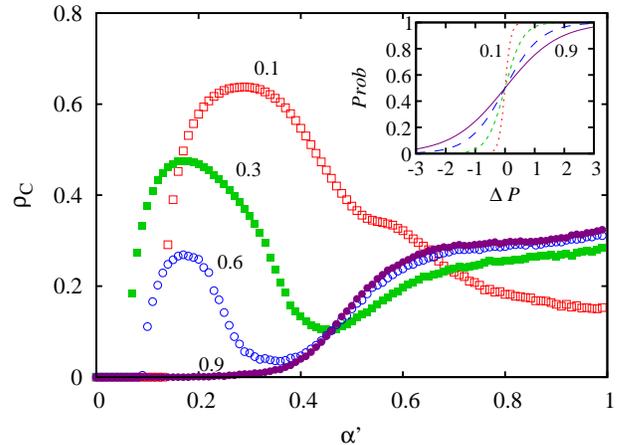,width=8.5cm}}
\caption{Optimal cooperation levels for square lattice at $T=1.1$ by using different $K$ values, as indicated. On the $x$-axis $\alpha^\prime = \alpha/K$ is a normalized parameter for proper comparison. The inset shows the related invasion probability functions dependent on payoff differences for the specific $K$ values. This function converges to a linear function as we increase $K$.}
\label{K} 
\end{figure}

As we argued, the reason why biased populations support cooperation is based on the fact that the intervention into the microscopic dynamics has asymmetric consequences on strategy invasions. Due to the strongly non-linear character of invasion probability, defined by Eq.~\ref{fermi}, a slight advantage of a higher temptation value results in a dramatic advantage for defectors that cannot be stated about cooperators whose payoff can hardly exceed a defector's value. This argument can be tested easily because if we apply a less non-linear invasion probability function then the  consequence of strategy-neutral intervention should be less biased, which would result in a mitigated cooperator supporting effect. Interestingly, a less non-linear probability function can be reached even in the framework of the used Fermi-function if we use higher noise values. This is illustrated in the inset of Fig.~\ref{K} where the originally step-like, strongly non-linear function tends to a linear function as we increase $K$. The main plot of Fig.~\ref{K} confirms our expectation because by increasing the noise value at a  fixed temptation the maximum value of the cooperation level decreases gradually. What is more, the cooperator supporting effect completely disappears above a critical noise value. Indeed, $\rho_C$ increases for high $\alpha$ values, but this is just a consequence of the artificial effect that too high $\alpha$ would result in a frozen state which may conserve the initial cooperation level.

\begin{figure}
\centerline{\epsfig{file=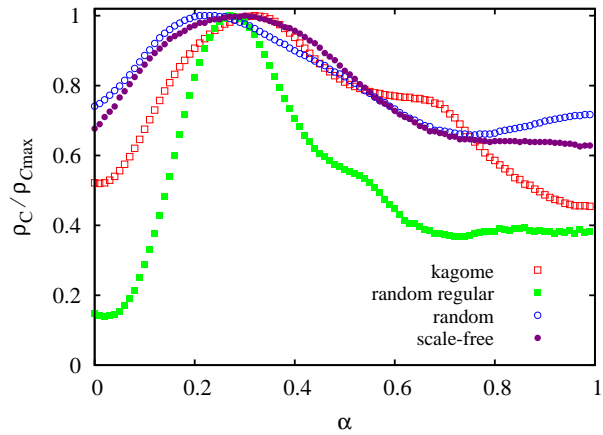,width=8.5cm}}
\caption{Fraction of cooperators in dependent on $\alpha$ for different interaction graphs as indicated. The applied temptation values are $T=1.09,$ 1.14, 1.06, and 1.02, respectively. For proper comparison all $\rho_C$ values are normalized with their maximal values. The system size of heterogeneous graphs are $N=5000$ where cooperation levels were averaged over 1000 independently generated configurations. For regular graphs we used $K=0.1$ while for heterogeneous graphs, where payoff values could be highly diverse, we applied $K=0.025$ to avoid the noise effect we discussed in the previous plot.}
\label{general} 
\end{figure}

Finally, we stress that our observations are not restricted to a square lattice interaction graph but can also be detected for other networks. Figure~\ref{general} illustrates that a positive impact at an intermediate $\alpha$ can be seen for other types of lattices, random graphs, and even for a highly heterogeneous scale-free network. The only compelling criterion for the population is to be structured where network reciprocity can work. Otherwise, in a well-mixed population, where interactions are just temporary, we cannot observe the stable coexistence of cooperator and defector strategies, hence the pattern formations discussed in Fig.~\ref{snapshots} are invalid.

The other crucial criterion is the non-linear payoff dependence of evolutionary success that is captured by the frequently used Fermi-type invasion probability, but other types of rules with similar features can also be cited here \cite{blume_l_geb93, szabo_pr07}. Indeed, the non-linear character of evolutionary dynamics is a broadly assumed and experimentally justified feature for a broad range of systems including biological, ecological, and economical examples \cite{aviles_eer99, archetti_jtb09, cadsby_jep07,  szolnoki_pre10, gonzalez-avella_pone11, chen_q_csf16, mas_jet16}.

\section{Discussion}

The evolution of cooperation is an intensively studied problem that has attracted hundreds of research papers proposing many sophisticated strategies and external conditions that could be helpful to overcome the original conflicts of individual and collective benefits \cite{pavlogiannis_srep15, xu_b_srep15, chen_ys_pa16, pei_zh_njp17, shen_c_pa17, amaral_pre18}. In the present work we have studied one of the simplest extensions of the basic prisoner's dilemma game and explored its possible consequences on the cooperation level.

It turns out that when the self-deception level of players is biased the general cooperation is elevated. Interestingly, to be overconfident, that is, to think more about their own achievement than it is worth in reality, is not just vital individually, but could also be useful collectively. Similarly, the presence of underconfident players can demolish the plausible advantage of defection. In both cases we can detect a dynamical effect that is responsible for this improvement. 

The underconfident attitude involves stimulated imitation of neighbors which results in general coordination of players. The emergence of locally homogeneous spots, however, directly supports cooperation strategy because it reveals the advantage of mutual cooperation. This mechanism can be identified in those systems where the inequality or heterogeneity of players were reported as a cooperator promoting circumstance. This heterogeneity may be originated from topological factors, like the difference between hub and periphery players, but could also be derived from individual differences. The latter could be strategy teaching or learning capacity \cite{szolnoki_epl07}, but also conformity \cite{xu_b_csf15, javarone_epl16} or the willingness to invest heterogeneously to different neighbors \cite{cao_xb_pa10, zhang_hf_pa12}.

Interestingly, overconfident players behave oppositely, they are reluctant to adopt neighboring strategies, still, the final outcome is very similar to those we observed for an underconfident society. In the latter case the aggressive propagation of successful defectors suffers more from the suppressed microscopic dynamics. In this way overconfidence attitude enhances the stability of evolving patterns hence a successful lonely defector cannot break the phalanx of cooperators even at a reasonably high temptation value. Put differently, the emergence of an overconfident attitude can enhance the network reciprocity that is already present in structured populations. We stress that the observed behavior is not only the spreading and final triumph of overconfident players over others with different attitude as reported in Ref.\cite{johnson_ddp_n11}, but the final outcome provides a higher well-being of the whole community via a higher cooperation level.

It is a common feature of both biased systems that a strategy-neutral intervention into the dynamics results in a highly biased impact on the evolution of strategies. This seemingly paradox behavior was also reported in systems where the cooperation level was sensitive to the applied dynamics \cite{szolnoki_pre09, szolnoki_pre14b}.
When the dynamic is suppressed, it retards the successful defector invasion more, while less successful cooperators benefit more from the stimulated imitations in the other case. In our present models these biased consequences are in close relation with the non-linear character of imitation dynamics that is a frequently observed phenomenon, which is broadly used in microscopic models. 

We hope that the present work gives insight into why overconfidence is a frequently emerging attitude that has a subtle impact on the success of the whole community.

\begin{acknowledgments}
This research was supported by the Hungarian National Research Fund (Grant K-120785) and by the National Natural Science Foundation of China (Grant No. 61503062).
\end{acknowledgments}

\end{document}